\documentclass[twocolumn]{aastex631}

\usepackage{graphicx}% Include figure files
\usepackage{bm}% bold math
\usepackage{amsmath}
\usepackage{amssymb}
\usepackage{amsthm}
\usepackage{multirow}
\usepackage{amsfonts}
\usepackage{color}
\usepackage{subfigure}
\usepackage{verbatim}
\shortauthors{Cang, Ma, Gao}
\newcommand{\fbh}{f_{\rm bh}} % pbh fraction
\newcommand{\mbh}{M_{\bullet}} % pbh mass
\newcommand{\ms}{{\rm M}_{\odot}} % solar mass
\newcommand{\sbh}{\sigma_{\bullet}} % pbh width
\newcommand{\neff}{N_{\rm{eff}}} % Neff
\newcommand{\dneff}{\Delta N_{\rm{eff}}} % delta Neff
\newcommand{\kbh}{k_{\bullet}}
\newcommand{\ps}{\mathcal{P}_{\mathcal{R}}}
\newcommand{\plk}{{\it Planck}}

\newcommand{\be}{\begin{equation}}
\newcommand{\ee}{\end{equation}}
\newcommand{\bea}{\begin{eqnarray}}
\newcommand{\eea}{\end{eqnarray}}

\DeclareRobustCommand{\Sec}[1]{Sec.~\ref{#1}}

\DeclareRobustCommand{\Fig}[1]{Fig.~\ref{#1}}

\DeclareRobustCommand{\Eq}[1]{Eq.~(\ref{#1})}

\DeclareRobustCommand{\r}[1]{{\rm #1}}

\begin{document}

\title{
Implications for primordial black holes 
from cosmological constraints on scalar-induced gravitational wave
}

\author{Junsong Cang}
\affiliation{Key Laboratory of Particle Astrophysics, Institute of High Energy Physics, Chinese Academy of Sciences, Beijing, 100049, China}
\affiliation{School of Physical Sciences, University of Chinese Academy of Sciences, Beijing, 100049, China}
\affiliation{Scuola Normale Superiore, Piazza dei Cavalieri 7, 56126 Pisa, Italy}
\author{Yin-Zhe Ma}
\thanks{Corresponding author: Y.-Z.~Ma, ma@ukzn.ac.za}
\affiliation{School of Chemistry and Physics, University of KwaZulu-Natal, Westville Campus, Private Bag X54001, Durban, 4000, South Africa}
\affiliation{NAOC–UKZN Computational Astrophysics Centre (NUCAC), University of KwaZulu-Natal, Durban, 4000, South Africa}
\affiliation{National Institute for Theoretical and Computational Sciences (NITheCS), South Africa}

\author{Yu Gao}
\thanks{Corresponding author: Y.~Gao, gaoyu@ihep.ac.cn}
\affiliation{Key Laboratory of Particle Astrophysics, Institute of High Energy Physics, Chinese Academy of Sciences, Beijing, 100049, China}

\begin{abstract}
Sufficiently large scalar perturbations in the early Universe can create over-dense regions that collapse into primordial black holes (PBH). 
This process is accompanied by the emission of scalar-induced gravitational waves (SIGW) that behave like an extra radiation component,
thus contributing to the relativistic degrees of freedom ($N_{\rm{eff}}$).
We show that the cosmological constraints on $N_{\rm{eff}}$ can be used to pose stringent limits on PBHs created from this particular scenario as well as the relevant small-scale curvature perturbation ($\mathcal{P}_{\mathcal{R}}(k)$).
We show that the combination of cosmic microwave background (CMB), 
baryon acoustic oscillation (BAO) and Big-Bang nucleosynthesis (BBN) datasets can exclude supermassive PBHs with peak mass $M_{\bullet} \in [5 \times 10^{5}, 5 \times 10^{10}]\,{\rm M}_{\odot}$ as the major component of dark matter,
while the detailed constraints depend on the shape of the PBHs mass distribution.
The future CMB mission like CMB-S4 can broaden this constraint window to a much larger range $M_{\bullet} \in [8 \times 10^{-5}, 5 \times 10^{10}]\,{\rm M}_{\odot}$, 
covering sub-stellar masses.
These limits on PBH correspond to a tightened constraint on $\mathcal{P}_{\mathcal{R}}$ on scales of $k \in [10, 10^{22}]\ {\rm{Mpc^{-1}}}$, 
much smaller than those probed by direct CMB and large-scale structure power spectra.
\end{abstract}


\section{Introduction} \label{sec:intro}
Large density fluctuation in the early Universe can lead to gravitational collapse of over-dense regions and form primordial black holes 
(PBHs;~\citealt{Hawking:1971ei,Carr:1974nx,Carr:1975qj}).
PBHs have been proposed to explain a range of observed black hole conundrums~\citep{Carr:2019kxo}, 
such as the existence of supermassive black holes~\citep{Bean:2002kx} and black hole merging events seen by LIGO and Virgo~\citep{LIGOScientific:2018mvr,Jedamzik:2020omx,Hutsi:2020sol}.
PBHs are also considered as a dark matter (DM) candidate~\citep{Carr:2016drx,Carr:2020xqk,Carr:2021bzv}
and their abundance has been constrained via various cosmological and astrophysical observations
~(\citealt{
Khlopov:2008qy,
Tashiro:2012qe,
Belotsky:2014kca,
Carr:2016drx,
Ali-Haimoud:2016mbv,
Wang:2016ana,
Clark:2016nst,
Clark:2018ghm,
Belotsky:2018wph,
Laha:2019ssq,
Chen:2019xse,
Carr:2020xqk,
Laha:2020ivk,
Ashoorioon:2019xqc,
Ashoorioon:2020hln,
Carr:2020gox,
Domenech:2020ssp,
Hutsi:2020sol,
Vaskonen:2020lbd,
Ray:2021mxu,
Yang:2021idt,
Domenech:2021wkk,
Carr:2021bzv,
Cang:2021owu,
Mittal:2021egv,
Zhou:2021tvp,
Auffinger:2022khh,
Ashoorioon:2022raz,
Karam:2022nym,
Wang:2022nml}, see~\citealt{Carr:2020xqk} for the review).

There are a plethora of mechanisms through which PBHs can be produced~\citep{Carr:2020gox},
such as the bubble collisions~\citep{Hawking:1982ga,Crawford:1982yz,La:1989st} and the collapse of cosmic strings~\citep{Hogan:1984zb,Hawking:1987bn,Polnarev:1988dh,Hansen:1999su}.
Here we focus on one particular scenario in which PBHs are formed from overdensities produced by an inflation-induced enhancement on primordial curvature perturbation
~\citep{
Carr:2009jm,
Inomata:2018epa,
Chen:2021nio,
Kimura:2021sqz,
Wang:2022nml}.
At large scales $k \lesssim 1 ~\r{Mpc^{-1}}$,
the curvature perturbation power spectrum $\ps$ has been precisely measured by cosmic microwave background (CMB) and large-scale structure (LSS) observations~\citep{Hunt:2015iua,Planck2018_parameters}, 
however the small-scale power is still poorly constrained
~\citep{
Inomata:2018epa,
Byrnes:2018txb,
Pi:2020otn,
Chen:2021nio,
Yuan:2021qgz,
Kimura:2021sqz,
Wang:2022nml}.
Thus, the enhanced small scale $\ps$ can potentially produce enough PBHs to explain {\it all} dark matter without violating the existing observational bounds.

Associated with PBH production,
enhanced $\ps$ inevitably generates scalar-induced gravitational wave (SIGW) at second order
~\citep{
Carbone:2004iv,
Nakamura:2004rm,
Vaskonen:2020lbd,
Yuan:2021qgz,
Chen:2021nio,
Karam:2022nym},
which can potentially be observed by gravitational wave (GW) detectors like Taiji~\citep{Hu:2017mde} and LISA~\citep{LISA:2017pwj}.
After horizon-crossing,
SIGW can be considered as dark radiation (DR;~\citealt{Chacko:2015noa,Serpico:2019bcj,Takahashi:2019ypv}) in that it free-streams with an energy density redshifting as $(1+z)^4$~\citep{Inomata:2018epa,Chen:2021nio}.
From a cosmological point of view,
the gravitational behavior of SIGW is indistinguishable from a relativistic species such as the massless neutrino. 
Therefore, 
just as other forms of dark radiation (e.g. primordial gravitational wave~\citep{Meerburg:2015zua,Aich:2019obd}, 
axion~\citep{Green:2019glg} and sterile neutrinos~\citep{Takahashi:2019ypv,Green:2019glg}), 
the energy density of SIGW DR can be parameterised by an additional effective degree of freedom 
\begin{eqnarray}
\dneff \equiv \neff - \neff^{\r{SM}}, \label{eq:def_Neff1}
\end{eqnarray}
where $\neff$ describes the total effects of relativistic species containing both neutrino and SIGW, 
and $\neff^{\r{SM}}=3.046$ is the prediction from the standard model (SM) of particle physics~\citep{Mangano:2001iu,Mangano:2005cc,deSalas:2016ztq}.

As a measure of cosmic radiation density,
a higher $\neff$ can delay radiation-to-matter equality and change the size of the sound horizon~\citep{Aich:2019obd}, 
which can leave distinctive features on CMB anisotropies~\citep{Hou:2011ec,Follin:2015hya},
baryon acoustic oscillations (BAO) and Big-Bang nucleosynthesis (BBN)~\citep{Wallisch:2018rzj}. 
Currently the leading $\dneff$ constraint is set by measuring the damping tail and the phase shift in the CMB anisotropy spectrum~\citep{Hou:2011ec,Planck:2015fie,Follin:2015hya,Wallisch:2018rzj,Aich:2019obd}, 
which reads $\dneff < 0.3$ at 95\% confidence level (C.L.) from the latest {\it Planck} CMB data~\citep{Planck2018_parameters}.
Future CMB missions are expected to give significant improvements on $\dneff$~\citep{CMB-S4:2016ple,CORE:2016npo,Alvarez:2019rhd}. 
For example, the CMB Stage IV (S4) experiment is aiming to constrain $\dneff < 0.027$ at 2$\sigma$ C.L.~\citep{Baumann:2015rya,CMB-S4:2016ple,Baumann:2016wac,Wallisch:2018rzj}.

Here we use the SIGW upper bounds inferred from cosmological constraints on $\dneff$ to place limits on both small-scale $\ps$ and the associated PBH abundance. 
The structure of this paper is as follows:
\Sec{PBH_model} reviews PBH production induced by enhanced curvature perturbation.
\Sec{SIGW_DR} discusses the energy density of SIGW and its connection with $\neff$.
 Our results are presented in \Sec{Results__} and we conclude in~\Sec{Summafdh_dshf}.
 
\section{PBH model}
\label{PBH_model}

In order for inflation to produce density fluctuation required for efficient PBH formation, 
the primordial curvature perturbation power spectrum $\ps$ needs to be boosted to $10^{-2}$ on small scales ($k\gtrsim 1\,{\rm Mpc}^{-1}$)
~\citep{
Josan:2009qn,
Bringmann:2011ut,
Garcia-Bellido:2017mdw,
Pi:2020otn,
Green:2020jor,
Chen:2021nio,
Karam:2022nym}, 
which can be realised in several inflation 
theories~\citep{
Kawasaki:1997ju,
Yokoyama:1998pt,
Kohri:2012yw,
Garcia-Bellido:2017mdw,
Kannike:2017bxn,
Cai:2019bmk,
Pi:2020otn}. 
As a good approximation for a wide range of curvature perturbations~\citep{Inomata:2018epa,Pi:2020otn,Chen:2021nio,Yuan:2021qgz,Domenech:2021ztg}, 
we adopt a log-normal $\ps$ model peaked at the scale $\kbh$ which was motivated in the Horndeski theory of gravity~\citep{Inomata:2018epa,Chen:2021nio}
\be
\ps
=
A_{\r{s}}
\left(
\frac{k}{k_*}
\right)^{n_{\r{s}}-1}
\left[
1+
\frac{A}
{\sqrt{2 \pi \sigma^2}}
\r{e}
^
{
-
\left(\ln\frac{k}{\kbh}
\right)^2
/2\sigma^2
}
\right],
\label{duyu8}
\ee
where $A_{\r{s}}=2.1\times 10^{-9}$ and $n_{\r{s}}=0.9665$ are the amplitude and spectral index of primordial fluctuations fixed at the pivot scale $k_{\ast}=0.05\ \r{Mpc}^{-1}$~\citep{Planck2018_parameters}.
Model parameters $A$, $\kbh$ and $\sigma$ describe the amplitude, location and width of the enhanced peak.
For large scales ($k \ll \kbh$) or for $A \to 0$,
$\ps$ is reduced to a power-law primordial spectrum. 

Once the large scalar fluctuations crossed the Hubble radius during radiation dominated era, 
the overdensity above a certain threshold ($\delta_{\rm c}\simeq 0.45$, see e.g.~\citealt{Musco:2012au,Harada:2013epa,Carr:2020xqk}\footnote{A simple analytic estimate would give $\delta_{\rm c} \sim 0.3$~\citep{1975ApJ...201....1C}, but our value was suggested from recent analytic and numerical analyses~\citep{Musco:2012au,Harada:2013epa,Carr:2020xqk}.}) can gravitationally collapse into a PBH with 
mass~\citep{Carr:2009jm,Nakama:2016gzw,Ozsoy:2018flq,Chen:2021nio},
\be
M
=
3.16 \times 10^{12}
\left(
\frac{\gamma}{0.2}
\right)
\left(
\frac{g_{\ast}(T)}{106.75}
\right)^{-1/6}
\left(
\frac{k}{\rm Mpc^{-1}}
\right)^{-2}
\ms,
\label{k2m_relation}
\ee
where $\gamma$ is the collapse efficiency,
for which we adopt a typical value of $\gamma = 0.2$
~\citep{
1975ApJ...201....1C,
Carr:2009jm,
Ozsoy:2018flq,
Chen:2021nio}.
$g_{\ast}(T)$ is the total number of effectively massless degrees of freedom (those species with $m\ll T$) at horizon-crossing ($k=aH$), which we adopted SM particle content (${\rm SU(3)}_{\rm C}\otimes {\rm SU(2)}_{\rm L}\otimes {\rm U(1)}_{\rm Y}$ theory~\citep{KolbTurner1990,Wallisch:2018rzj}).

We use Press-Schechter model~\citep{Press:1973iz} to calculate
the fraction of Universe's density collapsing into PBHs with mass $M$
~\citep{1975ApJ...201....1C,
Ozsoy:2018flq,
Carr:2020xqk,
Chen:2021nio}
\begin{eqnarray}
\beta (M)
& \equiv & \rho_{\bullet} / \rho_{\rm cr}(z) \nonumber \\
&=&
2
\int_{\delta_{\rm c}}^{\infty}
{\rm d} \delta
\frac{1}{\sqrt{2 \pi \bar{\sigma}^2}}
\exp
\left(
-
\frac{\delta^2}{2 \bar{\sigma}^2}
\right) \nonumber \\
& \simeq &
\sqrt{
\frac{2 \bar{\sigma}^2}
{\pi \delta_{\rm c}^2}
}
\r{exp}
\left(
-\frac{\delta_{\rm c}^2}{2 \bar{\sigma}^2}
\right),
\label{7865rt54321qw}
\end{eqnarray}
where $\rho_{\bullet}$ and $\rho_{\rm cr}(z)$ are the PBH density and the critical density of the Universe at the collapsing time.
In deriving the third equality in \Eq{7865rt54321qw} we used $\delta_{\rm c} > \bar{\sigma}$,
which remains valid for all scenarios we explored.
$\bar{\sigma}^2$ is the variance of density fluctuations at the scale $k^{-1}$
~\citep{Young:2014ana,Ozsoy:2018flq,Chen:2021nio}
\be
\bar{\sigma}^2
(M)
=
\frac{16}{81}
\int
\r{d}
\ln
k'\,
\left(
\frac{k'}{k}
\right)^4
\ps
(k')
W^2\left(\frac{k'}{k} \right),
\label{8765rfgrtf}
\ee
where $k(M)$ function is defined via \Eq{k2m_relation}. 
$W(x)$ is a window function, which we use $\exp(-x^2/2)$~\citep{Ozsoy:2018flq,Chen:2021nio}.

Since PBHs behave as matter,
$\rho_{\bullet}/\rho_{\r{cr}}$ grows inversely proportional to temperature until matter-radiation equality,
and the current distribution of PBH abundance in different masses can be estimated via 
(see also ~\citealt{Young:2014ana,Inomata:2017okj,Ozsoy:2018flq,Chen:2021nio})
\be
\begin{aligned}
\Phi
&\equiv
\frac{\r{d}\fbh}{{\rm d} \ln M}
\\
&=
0.28
\left(
\frac{\beta}
{10^{-8}}
\right)
\left(
\frac{\gamma}{0.2}
\right)^{3/2}
\left(
\frac{g_{\ast}}{106.75}
\right)^{-1/4}
\left(
\frac{M}{\ms}
\right)^{-1/2}
,
\label{9876tyuuyt}
\end{aligned}
\ee
where $\fbh \equiv \rho_{\bullet} / \rho_{\rm c}$ is the fraction of cold dark matter in form of PBHs.
For convenience, 
we describe the distribution calculated from \Eq{9876tyuuyt} with a log-normal parameterisation~\citep{
Dolgov:1992pu,
Green:2016xgy,
Carr:2017jsz,
Carr:2020xqk,
Cang:2021owu}
\be
\Phi
=
\fbh
\frac{1}{\sqrt{2 \pi} \sbh}
\exp
\left[
-
\frac{\r{ln}(M/\mbh)^2}
{2 \sbh^2}
\right]
,
\label{dshiuy65}
\ee
which provides an excellent fit to the actual profile.
Model parameters $\mbh$ and $\sbh$ describe peak PBH mass and distribution width respectively,
and their values are calculated using a least-square fitting method.

\section{SIGW as dark radiation}
\label{SIGW_DR}

In addition to forming PBHs,
scalar perturbations also change the radiation quadruple moment and generate GW at second order,
which carries an energy density $\rho_{\r{GW}}$ that redshifts as radiation after the horizon-crossing~\citep{Inomata:2018epa,Chen:2021nio},
$\rho_{\r{GW}} \propto (1+z)^4$.
At the present day,
the GW energy density parameter $\Omega_{\rm GW} \equiv \left(\rho_{\rm GW} / \rho_{\rm cr}\right)_{z=0}$
follows the distribution~\citep{Ando:2018qdb,Kohri:2018awv,Inomata:2018epa}
\be
\begin{aligned}
\Psi
& \equiv
\frac{{\rm d}\Omega_{\rm GW}}{{\rm d}\ln k}
\\
& = 
0.29~
\Omega_{\r{r}}
\left(
\frac{106.75}{g_{\ast}}
\right)^{1/3}
\\
&
\ \ \ \ 
\times
\int^\infty_0 dv \int^{1+v}_{|1-v|} du \left[ \frac{4v^2-(1-u^2+v^2)^2}{4u^2v^2}\right]^2
\\
&\ \ \ \  \times \left(\frac{u^2+v^2-3}{2 u v}\right)^4 F(u,v) \ps(kv)\ps(ku),
\end{aligned}
\label{eq98GWenergy01}
\ee
\be
\begin{aligned}
F(u,v)
&=
\left( \ln\left| \frac{3-(u+v)^2}{3-(u-v)^2}\right|-\frac{4 u v}{u^2+v^2-3}\right)^2
\\
&
+
\pi ^2 \Theta \left(u+v-\sqrt{3}\right)
,
\end{aligned}
\ee
where $\Omega_{\rm r}=9.1 \times 10^{-5}$ is present fractional radiation density assuming massless neutrinos.

After neutrino decoupling,
the cosmic radiation energy density ($\rho_{\r{r}}$) is a sum of CMB photon ($\gamma$), neutrino ($\nu$) and GW densities
\be
\rho_{\r{r}}
=
\rho_{\gamma}
+
\rho_{\nu}
+
\rho_{\r{GW}},
\label{sdfdiuyguih98}
\ee
where
\begin{eqnarray}
\rho_{\gamma}
&=&
\frac{\pi^2}{15}
T^4_{\gamma}, \nonumber \\
\rho_{\nu}
+
\rho_{\r{GW}}
&=&\frac{7 \pi^2}{120} N_{\r{eff}} 
T^4_{\nu}.
\label{dsvyuuhjt233}
\end{eqnarray}
Here $T_{\gamma}=2.728 (1+z)$\,K and $T_{\nu} = (4/11)^{1/3} T_{\gamma}$ are temperatures of CMB and neutrino respectively.
Because the behavior of GW density mimics that of neutrino,
\Eq{dsvyuuhjt233} counts the contribution of GW to the effective number of neutrino species as $\Delta N_{\rm eff}$ (Eq.~(\ref{eq:def_Neff1})). Comparing $\rho_{\r{GW}} = \Omega_{\r{GW}} \rho_{\r{cr}} (1+z)^4$ with \Eq{dsvyuuhjt233},
one can see that
\be
\Delta N_{\rm eff}
=8.3 \times 10^4 \,\Omega_{\r{GW}}(\vec{\theta}),
\label{dsvuytdui}
\ee
where
$\vec{\theta}$ indicates the model parameters. 
Depending on the choice of parameterisation,
$\vec{\theta}$ can be either the $\ps$ parameters $(A,\sigma,\kbh)$ defined in \Eq{duyu8},
or the PBH parameters $(\fbh,\sbh,\mbh)$ defined in \Eq{dshiuy65}.
For a given set of $\ps$ parameters,
the value of $\Omega_{\r{GW}}$ is directly determined by integrating Eq.~(\ref{eq98GWenergy01}). 
To obtain the relation between $\Omega_{\r{GW}}$ and PBH parameters,
we constructed a three dimensional grid in $\ps$ parameter space and calculated $\Omega_{\r{GW}}$ and $(\fbh,\sbh,\mbh)$ for each point on the grid,
using the large discrete sample of $\Omega_{\r{GW}} (\fbh,\sbh,\mbh)$ function obtained in the process,
we then built an interpolation function to calculate $\Omega_{\r{GW}}$ for any $(\fbh,\sbh,\mbh)$ parameter values in between.

To derive our PBH and $\ps$ constraints,
we use the $\neff$ limits from~\citet{Planck2018_parameters},
which gives $\neff = 3.04 \pm 0.22$ at $95\%$ C.L. for {\it Planck}+BAO+BBN data (hereafter PBB)\footnote{
Note that there are several updated $\neff$ constraints from more recent CMB datasets,
e.g. {\plk} and ACTPol jointly constrains $\neff = 2.74 \pm 0.17$ at 68\% C.L.~\citep{ACT:2020gnv},
which is lower than standard model $\neff = 3.046$ by $1.8\sigma$ C.L. This might be due to unaccounted systematics in the data. The SPT-3G and {\plk} data gives $\neff = 3.00 \pm 0.18$ at 68\% C.L~\citep{SPT-3G:2022hvq},
which has larger error than the PBB limits.}.
Using Gaussian statistics, 
this can be inverted to an 95\% C.L. upper bound of $\dneff < 0.175$.
We also study the prospective constraints from a future CMB Stage IV experiment,
which is expected to constrain $\dneff < 0.027$ at 95\% confidence level~\citep{Baumann:2015rya,CMB-S4:2016ple,Baumann:2016wac,Wallisch:2018rzj}. 
Combining the current and future constraints, we have
\be
\dneff <
\left\{
\begin{array}{l}
0.175\ \ 
\text{{\plk} + BAO +BBN (PBB)}
\\
0.027
\ \ 
\text{CMB Stage IV (S4)}.
\end{array}
\right.
\label{dsoiihivgu}
\ee

Through \Eq{dsvuytdui},
\Eq{dsoiihivgu} is inverted to the upper bounds on SIGW density at $95\%$ C.L.
\be
\Omega_{\r{GW}} 
<
\left\{
\begin{array}{l}
2.11 \times 10^{-6}\ \ \ \ \ 
\text{(PBB)}
\\
3.25 \times 10^{-7}
\ \ \ \ \ 
\text{(S4}).
\end{array}
\right.
\label{dhivgdsu}
\ee
In the following, 
we will substitute Eq.~(\ref{duyu8}) into Eq.~(\ref{eq98GWenergy01}) to calculate $\Omega_{\rm GW}$ or numerically using the $\Omega_{\r{GW}} (\fbh,\sbh,\mbh)$ interpolation function, 
and then use the upper bound in Eq.~(\ref{dhivgdsu}) to constrain the model parameters. 
We also notice that because of the positive correlation between $N_{\rm eff}$ and $H_0$, 
adopting a different value of $H_0$ may lead to the shift of posterior distribution of $N_{\rm eff}$. 
In recent years, 
measurements from the local distance ladder~\citep{Riess:2016jrr,Riess:2018uxu,Riess2022} give a higher value of $H_{0}$ than the CMB~\citep{Planck2018_parameters,ACT:2020gnv,SPT-3G:2022hvq} and BAO measurements~\citep{Verde2022}, 
which are inconsistent with each other at almost $5\sigma$ C.L.~\citep{Verde2019}. 
However, we notice that there are still a lot of debates on the potential systematics of Cepheids measurement~\citep{Freedman2020,Efstathiou2020,Efstathiou2021}, 
which requires more data in the future to clarify the Cepheids measurement. 
The uncertainty of $N_{\rm eff}$ associated with this on-going debates of anchor galaxy distance certainly exceeds the scope of this paper, 
but we notice the reader for this potential effect on $N_{\rm eff}$ measurement.

\section{Results}
\label{Results__}

\begin{figure*}[t]
\centering
\subfigbottomskip=-200pt
\subfigcapskip=-7pt
%\subfigure{\includegraphics[width=8.8cm]{PR_bounds.png}\includegraphics[width=8.5cm]{PR_bounds_2D.png}}
\subfigure{\includegraphics[width=8.5cm]{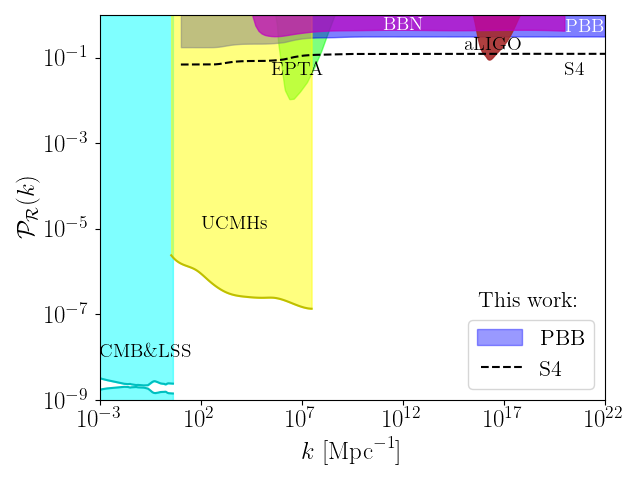}\includegraphics[width=8.3cm]{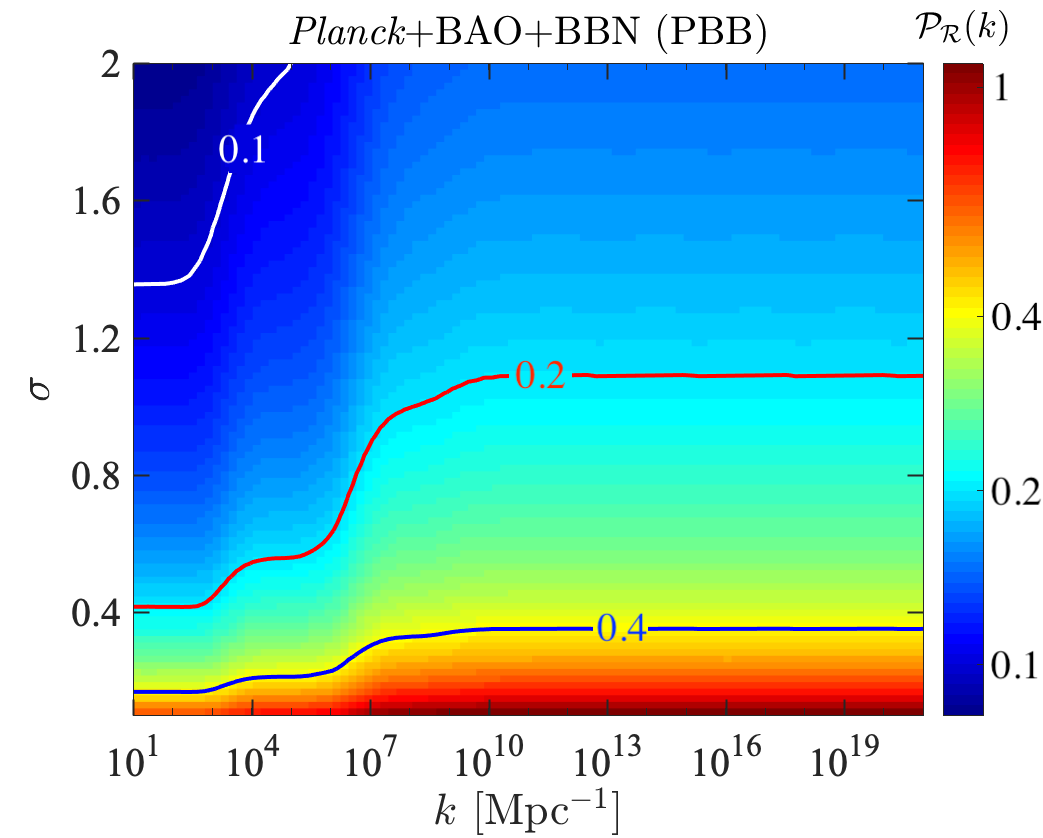}}
\caption{
{
%\footnotesize
\scriptsize
%\tiny
{\it Left}--
Comparison of $\ps$ constraints.
The shaded regions show the excluded parameter space. 
We present limits derived in this work for $\sigma = 0.5$ in blue shaded region and black dashed line,
corresponding to the current excluded space from PBB and the forecasted S4 upper bounds respectively.
Regions in cyan and yellow colours indicate the constraints from CMB and LSS~\citep{Hunt:2015iua}, 
and the non-detection of $\gamma$-rays from Ultracompact minihalos (UCMHs)~\citep{Bringmann:2011ut}. 
Regions in magenta, green, and brown colours show the exclusion of $\ps$ with width of $\sigma=0.5$, 
from Big-Bang Nucleosynthesis (BBN)~\citep{Kohri:2018awv,Inomata:2018epa}, European Pulsar Timing Array (EPTA)
and advanced LIGO (aLIGO)~\citep{Inomata:2018epa}. 
{\it Right}--
Our $\ps$ upper bounds for different $k$ and $\sigma$ parameters,
set by the current PBB datasets.
Prospective $\dneff$ limits from S4 can uniformly improve current PBB constraints by 60\%. 
}
}
\label{e2ftssdwu}
\end{figure*}

%\begin{figure}[htp]
\begin{figure}[b]
\centering
\includegraphics[width=8.5cm]{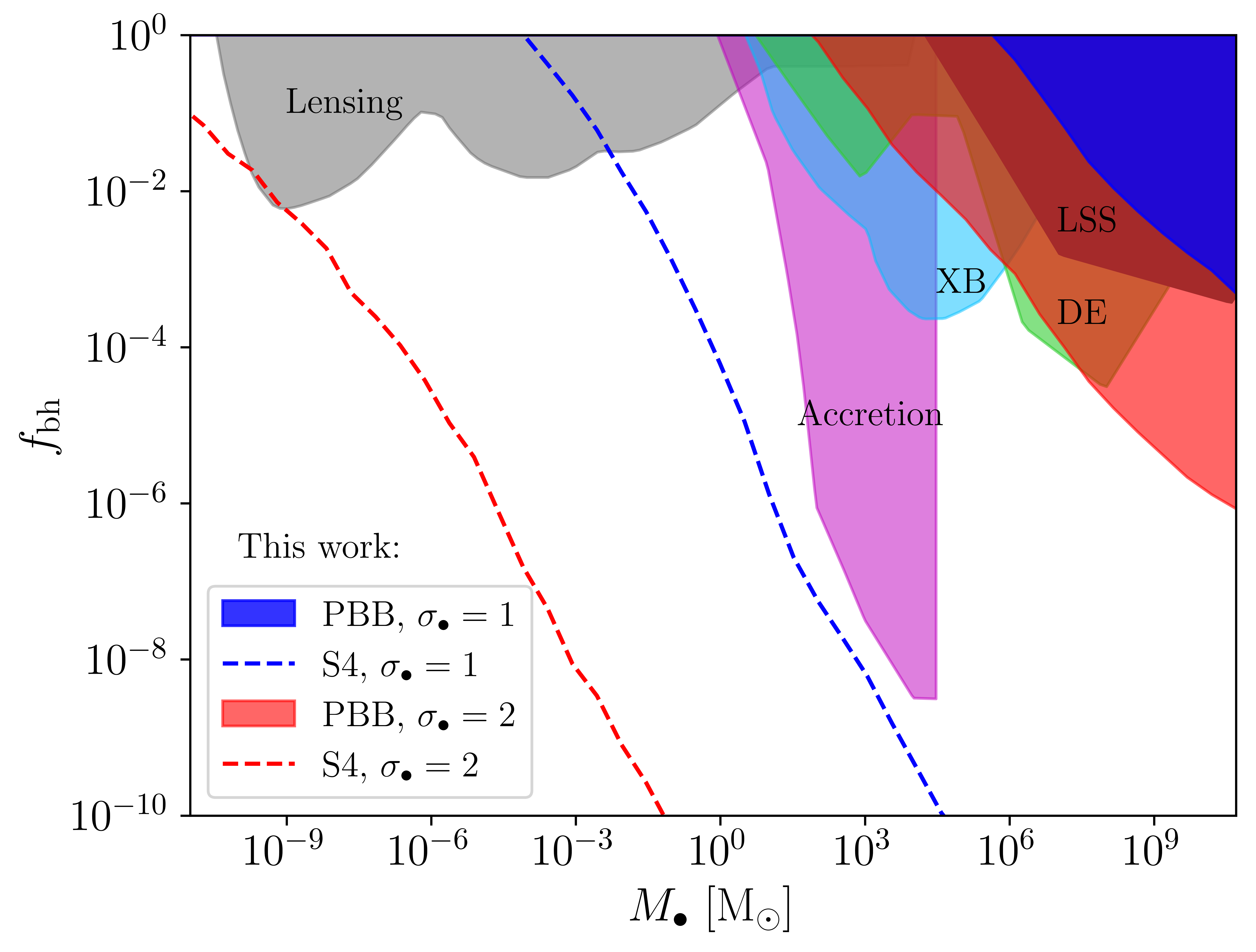}
\caption{
{
\scriptsize
%\footnotesize
Bounds on $\fbh$ as a function of PBH peak mass $\mbh$,
the filled regions indicate the excluded parameter space.
Our constraints are for PBHs produced by inflation-induced scalar perturbation,
with the blue and red regions showing the current PBB exclusion bounds on PBHs with distribution widths of 1 and 2 respectively,
whereas the blue ($\sigma_{\bullet} = 1$) and red ($\sigma_{\bullet} = 2$) dashed lines show the prospective limits from S4.
All other limits in this figure are collected from~\citet{Carr:2020xqk} and they apply to PBHs with monochromatic distribution (i.e. $\sigma_{\bullet} = 0$) irrespective of their formation mechanism,
from left to right, 
they are set by lensing (grey;~\citealt{EROS-2:2006ryy,Niikura:2017zjd}),
accretion (magenta;~\citealt{Serpico:2020ehh}), 
X-ray binary (XB, light blue;~\citealt{Inoue:2017csr}), 
dynamical effects (DE, green;~\citealt{1985ApJ...299..633L,Carr_1999,Brandt:2016aco,Carr:2020xqk}) and 
cosmological large-scale structure (LSS, brown;~\citealt{Carr:2018rid,Carr:2020xqk}). 
}
}
\label{e2ftdwu}
\end{figure}

For given $\sigma$ and $k$,
$\ps$ is determined by $\kbh$ and $A$ parameters. 
Once we set $\kbh=k \cdot \r{exp}[(1-n_{\r{s}}) \sigma ^2]$,
which gives a $\ps$ spectra that peaks at $k$ (see \Eq{duyu8}),
the upper bound on $\ps (k)$ can be obtained by fixing $A$ to its upper limit given by Eq.~(\ref{dhivgdsu}).
To avoid overlapping with existed large scales constraints ~\citep{Hunt:2015iua},
we restrict the solution to the range of $k > 10\ \r{Mpc}^{-1}$, which corresponds to $\mbh < 5 \times 10^{10}\,\ms$ from Eq.~(\ref{k2m_relation}). 

As shown in \Fig{e2ftssdwu},
in the majority of parameter space we explored,
our results show that PBB constrains $\ps$ to be $\sim \mathcal{O}(10^{-1})$.
The constraint becomes more stringent for a wider spectral width $\sigma$,
which is because the peak amplitude of $\ps$ is roughly inversely proportional to $\sigma$.
For a sharp spectra with $\sigma=0.1$,
our constraint yields $\ps \lesssim 1$,
whereas the limit tightens to $\ps \lesssim 0.1$ for $\sigma=2$.
In most cases,
the linear part in \Eq{duyu8} can be safely ignored,
so that $\ps \propto A$ and $\Omega_{\r{GW}} \propto A^2$,
therefore our $\ps$ upper limit is proportional to the maximally allowed $\sqrt{\Omega_{\r{GW}}}$. 
Therefore, compared to PBB, 
the CMB-S4 experiment will uniformly improve its $\ps$ constraint by 60 per cent on all scales.

The left panel of \Fig{e2ftssdwu} compares our PBB and S4 results with other leading $\ps$ constraints from different astrophysical data
~\citep{Bringmann:2011ut,Hunt:2015iua,Inomata:2018epa}.
At $k < 4\,\r{Mpc}^{-1}$,
$\ps$ is well measured by CMB anisotropy and LSS to $\sim 2 \times 10^{-9}$~\citep{Hunt:2015iua}.
Between $k=[4, 3 \times 10^7]\,{\rm Mpc}^{-1}$,
the non-detection of gamma rays from ultracompact minihalos (UCMHs) gives $\ps \lesssim 10^{-6}$~\citep{Bringmann:2011ut}.
All other limits shown in Fig.~\ref{e2ftssdwu} are for $\ps$ with a log-normal width of $\sigma=0.5$, 
set by the BBN and GW observations, e.g.
EPTA (European Pulsar Timing Array) and aLIGO (advanced LIGO)~\citep{Inomata:2018epa}.
Our PBB results constrain $\ps \lesssim 0.28$,
which is the strongest $\ps$ limit for $k \in [10^8,10^{22}] \ \r{Mpc}^{-1}$ up-to-date. 
Projected limits from S4 further tighten to $\ps \lesssim 0.11$.
Future GW detectors such as Taiji~\citep{Hu:2017mde}, TianQin~\citep{TianQin:2015yph}, LISA~\citep{LISA:2017pwj} and SKA
~\citep{Carilli:2004nx,Moore:2014lga,Janssen:2014dka}
can probe $k$ in between $10^5$ to $10^{14}\ \r{Mpc}^{-1}$~\citep{Chen:2021nio}, 
potentially improving our constraints further.

% -------- Fig:BH 2D --------
\begin{figure*}[tp] 
\centering
\subfigbottomskip=-200pt
\subfigcapskip=-7pt
\subfigure{\includegraphics[width=8.5cm]{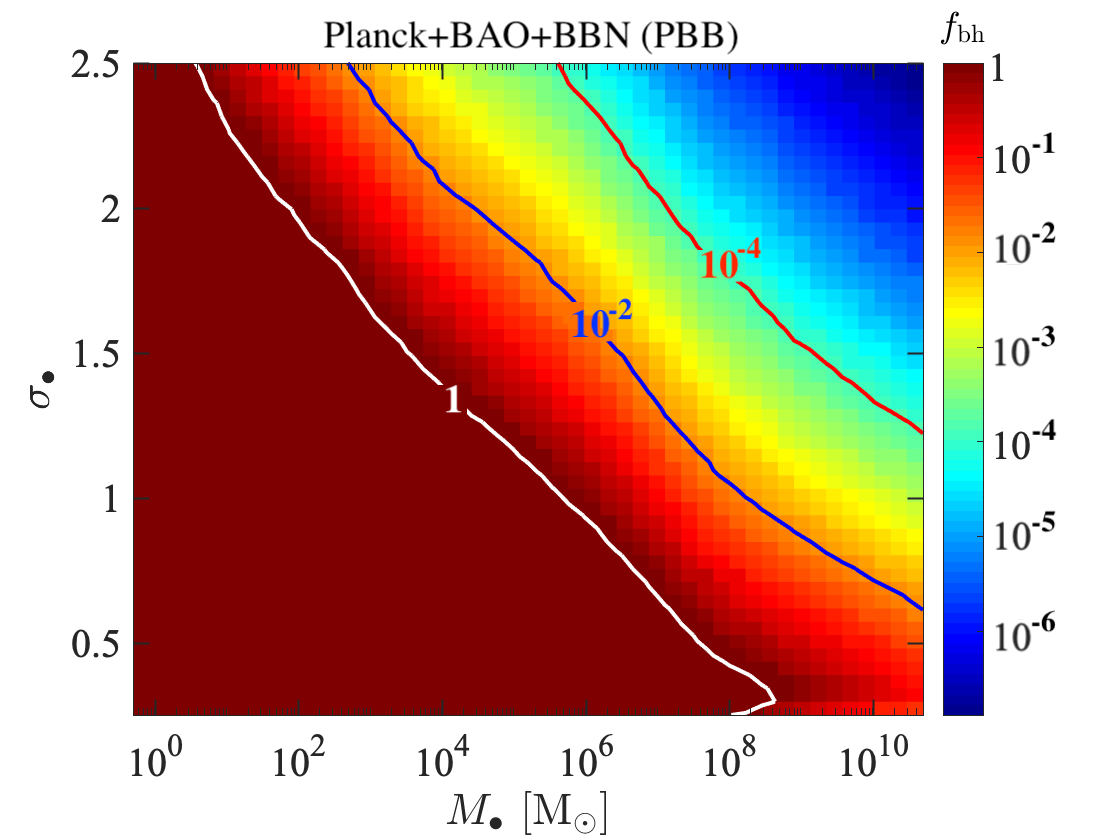}\includegraphics[width=8.5cm]{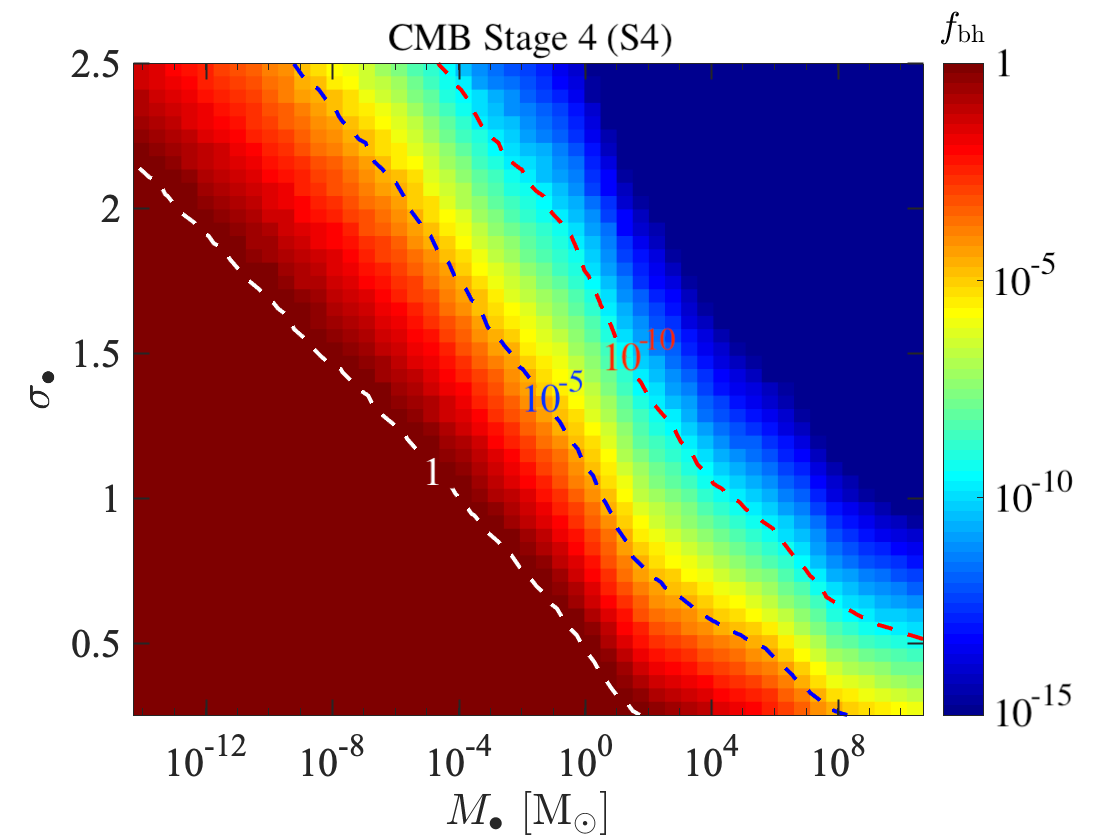}}
\caption{
{
\scriptsize
The 95\% C.L. upper bounds on $\fbh$ for different peak mass $M_{\bullet}$ and distribution width $\sigma_{\bullet}$,
applicable to PBHs formed from inflation-induced scalar perturbation.
The {\it left} and {\it right} panels show the current PBB constraints and the prospective S4 limits respectively.
The white contours indicate the parameter space where PBH can explain all dark matter ($\fbh = 1$).
}
}
\label{fiduy876}
\end{figure*}

As with the majority of PBH formation theories
~\citep{Dolgov:1992pu,Yokoyama:1998xd,Niemeyer:1999ak,Green:2016xgy,Carr:2017jsz,Bellomo:2017zsr,Pi:2017gih},
our PBHs follow an extended distribution and can be well described by the log-normal parameterisation in \Eq{dshiuy65}.
Using the $\Omega_{\r{GW}} (f_{\r{bh}},\sigma_{\r{bh}},M_{\r{bh}})$ interpolation function described in previous section,
it can be shown numerically that $\Omega_{\r{GW}}$ increases with $\fbh$,
and here we derive our $\fbh$ upper bounds by iteratively solving~\Eq{dhivgdsu}.
In \Fig{e2ftdwu} we show the $\fbh$ limits for PBHs with distribution widths of $\sbh = 1$ and $\sbh = 2$, 
along with the existing bounds on monochromatic PBH (assuming all PBHs having same mass) summarised in~\citet{Carr:2020xqk} and \citet{Serpico:2020ehh}.
Note that in addition to the inflation-induced scalar perturbation,
there are many other PBH formation scenarios~\citep{Carr:2020gox},
e.g. collapse of cosmic strings~\citep{Hogan:1984zb,Hawking:1987bn,Polnarev:1988dh,Hansen:1999su} and the bubble collisions~\citep{Hawking:1982ga,Crawford:1982yz,La:1989st},
thus caution should be taken while interpreting \Fig{e2ftdwu}. Our constraints applies to PBHs formed from enhanced scalar perturbation,
whereas the rest of the limits constrain PBHs regardless of their formation mechanism.

For constraint from PBB with $\sbh=1$,
the blue filled region in Fig.~\ref{e2ftdwu} shows that it excludes supermassive PBHs with 
$\mbh \in [5 \times 10^5,\ 5\times 10^{10}]\ \ms$ as the dominant DM component ($\fbh < 1$), 
spanning over 5 orders of magnitude. 
The limit covers the range set by X-ray binary (light blue; ~\citealt{Inoue:2017csr}) and LSS (brown;~\citealt{Carr:2020xqk}). 
For $\sbh=2$, 
the PBB exclusion window expands to $[10^2,\ 5\times 10^{10}]\ \ms$, 
which constitutes the widest PBH constraints up-to-date. 
The red and blue dashed lines show the sensitivity of projected CMB-S4 experiment on PBH abundance,
which can exclude PBHs in $[8 \times 10^{-5},\ 5\times 10^{10}]\ \ms$
while setting the most stringent PBH constraints in a large mass range of $[7 \times 10^{-3},\ 5\times 10^{10}]\ \ms$.
Compared to other leading constraints in the supermassive PBH window around $[3 \times 10^{4},\ 5\times 10^{10}]\ \ms$,
the limit from projected S4 can be stronger by more than 10 orders of magnitude.
\Fig{fiduy876} shows the complete constraints for a range of $\sbh$ values.
For fixed $\fbh$,
we find that $\Omega_{\r{GW}}$ increases with both $\sbh$ and $\mbh$,
therefore our $\fbh$ limit is tightened as we increase either $\sbh$ or $\mbh$.

\section{Summary}
\label{Summafdh_dshf}
Scalar-perturbation-induced PBH formation events emit SIGWs that behave like a relativistic species, 
so that the measured $\dneff$ can yield stringent limits on the scalar power spectra $\ps$ on small scales $10<k<10^{22}\ {\r{Mpc}}^{-1}$, much smaller than the direct CMB power spectra measurement. 
Using a log-normal parameterisation for $\ps$ arised from Horndeski gravity theory~\citep{Chen:2021nio} (also a good approximation for a wide class of perturbation theories), we show that {\plk} CMB data combined with BBN and BAO datasets (PBB) gives the currently most stringent $\ps$ constraints in $k \in [10^8,10^{22}] \ \r{Mpc}^{-1}$. 
For PBHs with a log-normal width of $\sbh=1$,
PBB exclude supermassive PBHs with peak mass $\mbh \in [5 \times 10^5,\ 5\times 10^{10}]\ \ms$ as the dominant DM component.
Future CMB-S4 can probe $\mbh \in [8 \times 10^{-5},\ 5\times 10^{10}]\ \ms$ mass window
and potentially improve $\ps$ limits of PBB by 60\%.
\\
\\
%\begin{acknowledgments}
J.C. and Y.G. acknowledges support from the National Natural Science Foundation of China (12275278), the China Scholarship Council (CSC, No.202104910395) and partially from the Ministry of Science and Technology of China (2020YFC2201601). Y.Z.M. is supported by the National Research Foundation of South Africa under grant No. 120385 and No. 120378, NITheCS program ``New Insights into Astrophysics and Cosmology with Theoretical Models confronting Observational Data'', and National Natural Science Foundation of China with project 12047503. J.C. thanks for the hospitality at the Cosmology Group of SNS during this project.
%\end{acknowledgments}

\vspace{5mm}
\bibliography{refs.bib}

\begin{thebibliography}{}
\expandafter\ifx\csname natexlab\endcsname\relax\def\natexlab#1{#1}\fi
\providecommand{\url}[1]{\href{#1}{#1}}
\providecommand{\dodoi}[1]{doi:~\href{http://doi.org/#1}{\nolinkurl{#1}}}
\providecommand{\doeprint}[1]{\href{http://ascl.net/#1}{\nolinkurl{http://ascl.net/#1}}}
\providecommand{\doarXiv}[1]{\href{https://arxiv.org/abs/#1}{\nolinkurl{https://arxiv.org/abs/#1}}}

\bibitem[{Abazajian {et~al.}(2016)}]{CMB-S4:2016ple}
Abazajian, K.~N., {et~al.} 2016, arXiv e-prints, arXiv:1610.02743.
\newblock \doarXiv{1610.02743}

\bibitem[{Abbott {et~al.}(2019)}]{LIGOScientific:2018mvr}
Abbott, B.~P., {et~al.} 2019, Phys. Rev. X, 9, 031040,
  \dodoi{10.1103/PhysRevX.9.031040}

\bibitem[{Ade {et~al.}(2016)}]{Planck:2015fie}
Ade, P. A.~R., {et~al.} 2016, Astron. Astrophys., 594, A13,
  \dodoi{10.1051/0004-6361/201525830}

\bibitem[{Aich {et~al.}(2020)Aich, Ma, Dai, \& Xia}]{Aich:2019obd}
Aich, M., Ma, Y.-Z., Dai, W.-M., \& Xia, J.-Q. 2020, Phys. Rev. D, 101, 063536,
  \dodoi{10.1103/PhysRevD.101.063536}

\bibitem[{Aiola {et~al.}(2020)}]{ACT:2020gnv}
Aiola, S., {et~al.} 2020, JCAP, 12, 047, \dodoi{10.1088/1475-7516/2020/12/047}

\bibitem[{{Ali-Ha{\"\i}moud} \& {Kamionkowski}(2017)}]{Ali-Haimoud:2016mbv}
{Ali-Ha{\"\i}moud}, Y., \& {Kamionkowski}, M. 2017, \prd, 95, 043534,
  \dodoi{10.1103/PhysRevD.95.043534}

\bibitem[{Amaro-Seoane {et~al.}(2017)}]{LISA:2017pwj}
Amaro-Seoane, P., {et~al.} 2017, arXiv e-prints, arXiv:1702.00786.
\newblock \doarXiv{1702.00786}

\bibitem[{Ando {et~al.}(2018)Ando, Inomata, \& Kawasaki}]{Ando:2018qdb}
Ando, K., Inomata, K., \& Kawasaki, M. 2018, Phys. Rev. D, 97, 103528,
  \dodoi{10.1103/PhysRevD.97.103528}

\bibitem[{Ashoorioon {et~al.}(2022)Ashoorioon, Rezazadeh, \&
  Rostami}]{Ashoorioon:2022raz}
Ashoorioon, A., Rezazadeh, K., \& Rostami, A. 2022, Phys. Lett. B, 835, 137542,
  \dodoi{10.1016/j.physletb.2022.137542}

\bibitem[{Ashoorioon {et~al.}(2021{\natexlab{a}})Ashoorioon, Rostami, \&
  Firouzjaee}]{Ashoorioon:2019xqc}
Ashoorioon, A., Rostami, A., \& Firouzjaee, J.~T. 2021{\natexlab{a}}, JHEP, 07,
  087, \dodoi{10.1007/JHEP07(2021)087}

\bibitem[{Ashoorioon {et~al.}(2021{\natexlab{b}})Ashoorioon, Rostami, \&
  Firouzjaee}]{Ashoorioon:2020hln}
---. 2021{\natexlab{b}}, Phys. Rev. D, 103, 123512,
  \dodoi{10.1103/PhysRevD.103.123512}

\bibitem[{{Auffinger}(2022)}]{Auffinger:2022khh}
{Auffinger}, J. 2022, arXiv e-prints, arXiv:2206.02672.
\newblock \doarXiv{2206.02672}

\bibitem[{Balkenhol {et~al.}(2022)}]{SPT-3G:2022hvq}
Balkenhol, L., {et~al.} 2022.
\newblock \doarXiv{2212.05642}

\bibitem[{Baumann {et~al.}(2016{\natexlab{a}})Baumann, Green, Meyers, \&
  Wallisch}]{Baumann:2015rya}
Baumann, D., Green, D., Meyers, J., \& Wallisch, B. 2016{\natexlab{a}}, JCAP,
  01, 007, \dodoi{10.1088/1475-7516/2016/01/007}

\bibitem[{Baumann {et~al.}(2016{\natexlab{b}})Baumann, Green, \&
  Wallisch}]{Baumann:2016wac}
Baumann, D., Green, D., \& Wallisch, B. 2016{\natexlab{b}}, Phys. Rev. Lett.,
  117, 171301, \dodoi{10.1103/PhysRevLett.117.171301}

\bibitem[{Bean \& Magueijo(2002)}]{Bean:2002kx}
Bean, R., \& Magueijo, J. 2002, Phys. Rev. D, 66, 063505,
  \dodoi{10.1103/PhysRevD.66.063505}

\bibitem[{Bellomo {et~al.}(2018)Bellomo, Bernal, Raccanelli, \&
  Verde}]{Bellomo:2017zsr}
Bellomo, N., Bernal, J.~L., Raccanelli, A., \& Verde, L. 2018, JCAP, 01, 004,
  \dodoi{10.1088/1475-7516/2018/01/004}

\bibitem[{Belotsky {et~al.}(2014)Belotsky, Dmitriev, Esipova, Gani, Grobov,
  Khlopov, Kirillov, Rubin, \& Svadkovsky}]{Belotsky:2014kca}
Belotsky, K.~M., Dmitriev, A.~D., Esipova, E.~A., {et~al.} 2014, Mod. Phys.
  Lett. A, 29, 1440005, \dodoi{10.1142/S0217732314400057}

\bibitem[{Belotsky {et~al.}(2019)Belotsky, Dokuchaev, Eroshenko, Esipova,
  Khlopov, Khromykh, Kirillov, Nikulin, Rubin, \&
  Svadkovsky}]{Belotsky:2018wph}
Belotsky, K.~M., Dokuchaev, V.~I., Eroshenko, Y.~N., {et~al.} 2019, Eur. Phys.
  J. C, 79, 246, \dodoi{10.1140/epjc/s10052-019-6741-4}

\bibitem[{Brandt(2016)}]{Brandt:2016aco}
Brandt, T.~D. 2016, Astrophys. J. Lett., 824, L31,
  \dodoi{10.3847/2041-8205/824/2/L31}

\bibitem[{Bringmann {et~al.}(2012)Bringmann, Scott, \&
  Akrami}]{Bringmann:2011ut}
Bringmann, T., Scott, P., \& Akrami, Y. 2012, Phys. Rev. D, 85, 125027,
  \dodoi{10.1103/PhysRevD.85.125027}

\bibitem[{Byrnes {et~al.}(2019)Byrnes, Cole, \& Patil}]{Byrnes:2018txb}
Byrnes, C.~T., Cole, P.~S., \& Patil, S.~P. 2019, JCAP, 06, 028,
  \dodoi{10.1088/1475-7516/2019/06/028}

\bibitem[{Cai {et~al.}(2020)Cai, Guo, Liu, Liu, \& Yang}]{Cai:2019bmk}
Cai, R.-G., Guo, Z.-K., Liu, J., Liu, L., \& Yang, X.-Y. 2020, JCAP, 06, 013,
  \dodoi{10.1088/1475-7516/2020/06/013}

\bibitem[{Cang {et~al.}(2022)Cang, Gao, \& Ma}]{Cang:2021owu}
Cang, J., Gao, Y., \& Ma, Y.-Z. 2022, JCAP, 03, 012,
  \dodoi{10.1088/1475-7516/2022/03/012}

\bibitem[{Carbone \& Matarrese(2005)}]{Carbone:2004iv}
Carbone, C., \& Matarrese, S. 2005, Phys. Rev. D, 71, 043508,
  \dodoi{10.1103/PhysRevD.71.043508}

\bibitem[{Carilli \& Rawlings(2004)}]{Carilli:2004nx}
Carilli, C.~L., \& Rawlings, S. 2004, New Astron. Rev., 48, 979,
  \dodoi{10.1016/j.newar.2004.09.001}

\bibitem[{Carr {et~al.}(2021{\natexlab{a}})Carr, Clesse, Garc\'\i{}a-Bellido,
  \& K\"uhnel}]{Carr:2019kxo}
Carr, B., Clesse, S., Garc\'\i{}a-Bellido, J., \& K\"uhnel, F.
  2021{\natexlab{a}}, Phys. Dark Univ., 31, 100755,
  \dodoi{10.1016/j.dark.2020.100755}

\bibitem[{Carr {et~al.}(2021{\natexlab{b}})Carr, Kohri, Sendouda, \&
  Yokoyama}]{Carr:2020gox}
Carr, B., Kohri, K., Sendouda, Y., \& Yokoyama, J. 2021{\natexlab{b}}, Rept.
  Prog. Phys., 84, 116902, \dodoi{10.1088/1361-6633/ac1e31}

\bibitem[{Carr \& Kuhnel(2020)}]{Carr:2020xqk}
Carr, B., \& Kuhnel, F. 2020, Ann. Rev. Nucl. Part. Sci., 70, 355,
  \dodoi{10.1146/annurev-nucl-050520-125911}

\bibitem[{Carr \& Kuhnel(2022)}]{Carr:2021bzv}
---. 2022, SciPost Phys. Lect. Notes, 48, 1,
  \dodoi{10.21468/SciPostPhysLectNotes.48}

\bibitem[{Carr {et~al.}(2016)Carr, Kuhnel, \& Sandstad}]{Carr:2016drx}
Carr, B., Kuhnel, F., \& Sandstad, M. 2016, Phys. Rev. D, 94, 083504,
  \dodoi{10.1103/PhysRevD.94.083504}

\bibitem[{Carr {et~al.}(2017)Carr, Raidal, Tenkanen, Vaskonen, \&
  Veerm\"ae}]{Carr:2017jsz}
Carr, B., Raidal, M., Tenkanen, T., Vaskonen, V., \& Veerm\"ae, H. 2017, Phys.
  Rev. D, 96, 023514, \dodoi{10.1103/PhysRevD.96.023514}

\bibitem[{Carr \& Silk(2018)}]{Carr:2018rid}
Carr, B., \& Silk, J. 2018, Mon. Not. Roy. Astron. Soc., 478, 3756,
  \dodoi{10.1093/mnras/sty1204}

\bibitem[{Carr(1975)}]{Carr:1975qj}
Carr, B.~J. 1975, Astrophys. J., 201, 1, \dodoi{10.1086/153853}

\bibitem[{{Carr}(1975)}]{1975ApJ...201....1C}
{Carr}, B.~J. 1975, \apj, 201, 1, \dodoi{10.1086/153853}

\bibitem[{Carr \& Hawking(1974)}]{Carr:1974nx}
Carr, B.~J., \& Hawking, S.~W. 1974, Mon. Not. Roy. Astron. Soc., 168, 399

\bibitem[{Carr {et~al.}(2010)Carr, Kohri, Sendouda, \& Yokoyama}]{Carr:2009jm}
Carr, B.~J., Kohri, K., Sendouda, Y., \& Yokoyama, J. 2010, Phys. Rev. D, 81,
  104019, \dodoi{10.1103/PhysRevD.81.104019}

\bibitem[{Carr \& Sakellariadou(1999)}]{Carr_1999}
Carr, B.~J., \& Sakellariadou, M. 1999, The Astrophysical Journal, 516, 195,
  \dodoi{10.1086/307071}

\bibitem[{Chacko {et~al.}(2015)Chacko, Cui, Hong, \& Okui}]{Chacko:2015noa}
Chacko, Z., Cui, Y., Hong, S., \& Okui, T. 2015, Phys. Rev. D, 92, 055033,
  \dodoi{10.1103/PhysRevD.92.055033}

\bibitem[{{Chen} {et~al.}(2021){Chen}, {Koh}, \& {Tumurtushaa}}]{Chen:2021nio}
{Chen}, P., {Koh}, S., \& {Tumurtushaa}, G. 2021, arXiv e-prints,
  arXiv:2107.08638.
\newblock \doarXiv{2107.08638}

\bibitem[{Chen {et~al.}(2020)Chen, Yuan, \& Huang}]{Chen:2019xse}
Chen, Z.-C., Yuan, C., \& Huang, Q.-G. 2020, Phys. Rev. Lett., 124, 251101,
  \dodoi{10.1103/PhysRevLett.124.251101}

\bibitem[{Clark {et~al.}(2018)Clark, Dutta, Gao, Ma, \&
  Strigari}]{Clark:2018ghm}
Clark, S., Dutta, B., Gao, Y., Ma, Y.-Z., \& Strigari, L.~E. 2018, Phys. Rev.
  D, 98, 043006, \dodoi{10.1103/PhysRevD.98.043006}

\bibitem[{Clark {et~al.}(2017)Clark, Dutta, Gao, Strigari, \&
  Watson}]{Clark:2016nst}
Clark, S., Dutta, B., Gao, Y., Strigari, L.~E., \& Watson, S. 2017, Phys. Rev.
  D, 95, 083006, \dodoi{10.1103/PhysRevD.95.083006}

\bibitem[{Crawford \& Schramm(1982)}]{Crawford:1982yz}
Crawford, M., \& Schramm, D.~N. 1982, Nature, 298, 538,
  \dodoi{10.1038/298538a0}

\bibitem[{de~Salas \& Pastor(2016)}]{deSalas:2016ztq}
de~Salas, P.~F., \& Pastor, S. 2016, JCAP, 07, 051,
  \dodoi{10.1088/1475-7516/2016/07/051}

\bibitem[{Di~Valentino {et~al.}(2018)}]{CORE:2016npo}
Di~Valentino, E., {et~al.} 2018, JCAP, 04, 017,
  \dodoi{10.1088/1475-7516/2018/04/017}

\bibitem[{Dolgov \& Silk(1993)}]{Dolgov:1992pu}
Dolgov, A., \& Silk, J. 1993, Phys. Rev. D, 47, 4244,
  \dodoi{10.1103/PhysRevD.47.4244}

\bibitem[{Dom\`enech(2021)}]{Domenech:2021ztg}
Dom\`enech, G. 2021, Universe, 7, 398, \dodoi{10.3390/universe7110398}

\bibitem[{Dom\`enech {et~al.}(2021{\natexlab{a}})Dom\`enech, Lin, \&
  Sasaki}]{Domenech:2020ssp}
Dom\`enech, G., Lin, C., \& Sasaki, M. 2021{\natexlab{a}}, JCAP, 04, 062,
  \dodoi{10.1088/1475-7516/2021/11/E01}

\bibitem[{Dom\`enech {et~al.}(2021{\natexlab{b}})Dom\`enech, Takhistov, \&
  Sasaki}]{Domenech:2021wkk}
Dom\`enech, G., Takhistov, V., \& Sasaki, M. 2021{\natexlab{b}}, Phys. Lett. B,
  823, 136722, \dodoi{10.1016/j.physletb.2021.136722}

\bibitem[{{Efstathiou}(2020)}]{Efstathiou2020}
{Efstathiou}, G. 2020, arXiv e-prints, arXiv:2007.10716,
  \dodoi{10.48550/arXiv.2007.10716}

\bibitem[{{Efstathiou}(2021)}]{Efstathiou2021}
---. 2021, \mnras, 505, 3866, \dodoi{10.1093/mnras/stab1588}

\bibitem[{Follin {et~al.}(2015)Follin, Knox, Millea, \& Pan}]{Follin:2015hya}
Follin, B., Knox, L., Millea, M., \& Pan, Z. 2015, Phys. Rev. Lett., 115,
  091301, \dodoi{10.1103/PhysRevLett.115.091301}

\bibitem[{{Freedman} {et~al.}(2020){Freedman}, {Madore}, {Hoyt}, {Jang},
  {Beaton}, {Lee}, {Monson}, {Neeley}, \& {Rich}}]{Freedman2020}
{Freedman}, W.~L., {Madore}, B.~F., {Hoyt}, T., {et~al.} 2020, \apj, 891, 57,
  \dodoi{10.3847/1538-4357/ab7339}

\bibitem[{Garcia-Bellido \& Ruiz~Morales(2017)}]{Garcia-Bellido:2017mdw}
Garcia-Bellido, J., \& Ruiz~Morales, E. 2017, Phys. Dark Univ., 18, 47,
  \dodoi{10.1016/j.dark.2017.09.007}

\bibitem[{Green(2016)}]{Green:2016xgy}
Green, A.~M. 2016, Phys. Rev. D, 94, 063530, \dodoi{10.1103/PhysRevD.94.063530}

\bibitem[{Green \& Kavanagh(2021)}]{Green:2020jor}
Green, A.~M., \& Kavanagh, B.~J. 2021, J. Phys. G, 48, 043001,
  \dodoi{10.1088/1361-6471/abc534}

\bibitem[{Green {et~al.}(2019)}]{Green:2019glg}
Green, D., {et~al.} 2019, Bull. Am. Astron. Soc., 51, 159.
\newblock \doarXiv{1903.04763}

\bibitem[{Hanany {et~al.}(2019)}]{Alvarez:2019rhd}
Hanany, S., {et~al.} 2019, in Bulletin of the American Astronomical Society,
  Vol.~51, 194.
\newblock \doarXiv{1908.07495}

\bibitem[{Hansen {et~al.}(2000)Hansen, Christensen, \& Larsen}]{Hansen:1999su}
Hansen, R.~N., Christensen, M., \& Larsen, A.~L. 2000, Int. J. Mod. Phys. A,
  15, 4433, \dodoi{10.1142/S0217751X00001450}

\bibitem[{{Harada} {et~al.}(2013){Harada}, {Yoo}, \& {Kohri}}]{Harada:2013epa}
{Harada}, T., {Yoo}, C.-M., \& {Kohri}, K. 2013, \prd, 88, 084051,
  \dodoi{10.1103/PhysRevD.88.084051}

\bibitem[{Hawking(1971)}]{Hawking:1971ei}
Hawking, S. 1971, Mon. Not. Roy. Astron. Soc., 152, 75

\bibitem[{Hawking(1989)}]{Hawking:1987bn}
Hawking, S.~W. 1989, Phys. Lett. B, 231, 237,
  \dodoi{10.1016/0370-2693(89)90206-2}

\bibitem[{Hawking {et~al.}(1982)Hawking, Moss, \& Stewart}]{Hawking:1982ga}
Hawking, S.~W., Moss, I.~G., \& Stewart, J.~M. 1982, Phys. Rev. D, 26, 2681,
  \dodoi{10.1103/PhysRevD.26.2681}

\bibitem[{Hogan(1984)}]{Hogan:1984zb}
Hogan, C.~J. 1984, Phys. Lett. B, 143, 87, \dodoi{10.1016/0370-2693(84)90810-4}

\bibitem[{Hou {et~al.}(2013)Hou, Keisler, Knox, Millea, \&
  Reichardt}]{Hou:2011ec}
Hou, Z., Keisler, R., Knox, L., Millea, M., \& Reichardt, C. 2013, Phys. Rev.
  D, 87, 083008, \dodoi{10.1103/PhysRevD.87.083008}

\bibitem[{Hu \& Wu(2017)}]{Hu:2017mde}
Hu, W.-R., \& Wu, Y.-L. 2017, Natl. Sci. Rev., 4, 685,
  \dodoi{10.1093/nsr/nwx116}

\bibitem[{Hunt \& Sarkar(2015)}]{Hunt:2015iua}
Hunt, P., \& Sarkar, S. 2015, JCAP, 12, 052,
  \dodoi{10.1088/1475-7516/2015/12/052}

\bibitem[{H\"utsi {et~al.}(2021)H\"utsi, Raidal, Vaskonen, \&
  Veerm\"ae}]{Hutsi:2020sol}
H\"utsi, G., Raidal, M., Vaskonen, V., \& Veerm\"ae, H. 2021, JCAP, 03, 068,
  \dodoi{10.1088/1475-7516/2021/03/068}

\bibitem[{Inomata {et~al.}(2017)Inomata, Kawasaki, Mukaida, Tada, \&
  Yanagida}]{Inomata:2017okj}
Inomata, K., Kawasaki, M., Mukaida, K., Tada, Y., \& Yanagida, T.~T. 2017,
  Phys. Rev. D, 96, 043504, \dodoi{10.1103/PhysRevD.96.043504}

\bibitem[{Inomata \& Nakama(2019)}]{Inomata:2018epa}
Inomata, K., \& Nakama, T. 2019, Phys. Rev. D, 99, 043511,
  \dodoi{10.1103/PhysRevD.99.043511}

\bibitem[{Inoue \& Kusenko(2017)}]{Inoue:2017csr}
Inoue, Y., \& Kusenko, A. 2017, JCAP, 10, 034,
  \dodoi{10.1088/1475-7516/2017/10/034}

\bibitem[{Janssen {et~al.}(2015)}]{Janssen:2014dka}
Janssen, G., {et~al.} 2015, PoS, AASKA14, 037, \dodoi{10.22323/1.215.0037}

\bibitem[{Jedamzik(2021)}]{Jedamzik:2020omx}
Jedamzik, K. 2021, Phys. Rev. Lett., 126, 051302,
  \dodoi{10.1103/PhysRevLett.126.051302}

\bibitem[{Josan {et~al.}(2009)Josan, Green, \& Malik}]{Josan:2009qn}
Josan, A.~S., Green, A.~M., \& Malik, K.~A. 2009, Phys. Rev. D, 79, 103520,
  \dodoi{10.1103/PhysRevD.79.103520}

\bibitem[{Kannike {et~al.}(2017)Kannike, Marzola, Raidal, \&
  Veerm\"ae}]{Kannike:2017bxn}
Kannike, K., Marzola, L., Raidal, M., \& Veerm\"ae, H. 2017, JCAP, 09, 020,
  \dodoi{10.1088/1475-7516/2017/09/020}

\bibitem[{{Karam} {et~al.}(2022){Karam}, {Koivunen}, {Tomberg}, {Vaskonen}, \&
  {Veerm{\"a}e}}]{Karam:2022nym}
{Karam}, A., {Koivunen}, N., {Tomberg}, E., {Vaskonen}, V., \& {Veerm{\"a}e},
  H. 2022, arXiv e-prints, arXiv:2205.13540.
\newblock \doarXiv{2205.13540}

\bibitem[{Kawasaki {et~al.}(1998)Kawasaki, Sugiyama, \&
  Yanagida}]{Kawasaki:1997ju}
Kawasaki, M., Sugiyama, N., \& Yanagida, T. 1998, Phys. Rev. D, 57, 6050,
  \dodoi{10.1103/PhysRevD.57.6050}

\bibitem[{Khlopov(2010)}]{Khlopov:2008qy}
Khlopov, M.~Y. 2010, Res. Astron. Astrophys., 10, 495,
  \dodoi{10.1088/1674-4527/10/6/001}

\bibitem[{Kimura {et~al.}(2021)Kimura, Suyama, Yamaguchi, \&
  Zhang}]{Kimura:2021sqz}
Kimura, R., Suyama, T., Yamaguchi, M., \& Zhang, Y.-L. 2021, JCAP, 04, 031,
  \dodoi{10.1088/1475-7516/2021/04/031}

\bibitem[{Kohri {et~al.}(2013)Kohri, Lin, \& Matsuda}]{Kohri:2012yw}
Kohri, K., Lin, C.-M., \& Matsuda, T. 2013, Phys. Rev. D, 87, 103527,
  \dodoi{10.1103/PhysRevD.87.103527}

\bibitem[{Kohri \& Terada(2018)}]{Kohri:2018awv}
Kohri, K., \& Terada, T. 2018, Phys. Rev. D, 97, 123532,
  \dodoi{10.1103/PhysRevD.97.123532}

\bibitem[{{Kolb} \& {Turner}(1990)}]{KolbTurner1990}
{Kolb}, E.~W., \& {Turner}, M.~S. 1990, {The early universe}, Vol.~69 (CRC
  press)

\bibitem[{La \& Steinhardt(1989)}]{La:1989st}
La, D., \& Steinhardt, P.~J. 1989, Phys. Lett. B, 220, 375,
  \dodoi{10.1016/0370-2693(89)90890-3}

\bibitem[{{Lacey} \& {Ostriker}(1985)}]{1985ApJ...299..633L}
{Lacey}, C.~G., \& {Ostriker}, J.~P. 1985, \apj, 299, 633,
  \dodoi{10.1086/163729}

\bibitem[{Laha(2019)}]{Laha:2019ssq}
Laha, R. 2019, Phys. Rev. Lett., 123, 251101,
  \dodoi{10.1103/PhysRevLett.123.251101}

\bibitem[{Laha {et~al.}(2020)Laha, Mu\~noz, \& Slatyer}]{Laha:2020ivk}
Laha, R., Mu\~noz, J.~B., \& Slatyer, T.~R. 2020, Phys. Rev. D, 101, 123514,
  \dodoi{10.1103/PhysRevD.101.123514}

\bibitem[{Luo {et~al.}(2016)}]{TianQin:2015yph}
Luo, J., {et~al.} 2016, Class. Quant. Grav., 33, 035010,
  \dodoi{10.1088/0264-9381/33/3/035010}

\bibitem[{Mangano {et~al.}(2002)Mangano, Miele, Pastor, \&
  Peloso}]{Mangano:2001iu}
Mangano, G., Miele, G., Pastor, S., \& Peloso, M. 2002, Phys. Lett. B, 534, 8,
  \dodoi{10.1016/S0370-2693(02)01622-2}

\bibitem[{Mangano {et~al.}(2005)Mangano, Miele, Pastor, Pinto, Pisanti, \&
  Serpico}]{Mangano:2005cc}
Mangano, G., Miele, G., Pastor, S., {et~al.} 2005, Nucl. Phys. B, 729, 221,
  \dodoi{10.1016/j.nuclphysb.2005.09.041}

\bibitem[{{Meerburg} {et~al.}(2015){Meerburg}, {Hlo{\v{z}}ek}, {Hadzhiyska}, \&
  {Meyers}}]{Meerburg:2015zua}
{Meerburg}, P.~D., {Hlo{\v{z}}ek}, R., {Hadzhiyska}, B., \& {Meyers}, J. 2015,
  \prd, 91, 103505, \dodoi{10.1103/PhysRevD.91.103505}

\bibitem[{Mittal {et~al.}(2022)Mittal, Ray, Kulkarni, \&
  Dasgupta}]{Mittal:2021egv}
Mittal, S., Ray, A., Kulkarni, G., \& Dasgupta, B. 2022, JCAP, 03, 030,
  \dodoi{10.1088/1475-7516/2022/03/030}

\bibitem[{Moore {et~al.}(2015)Moore, Cole, \& Berry}]{Moore:2014lga}
Moore, C.~J., Cole, R.~H., \& Berry, C. P.~L. 2015, Class. Quant. Grav., 32,
  015014, \dodoi{10.1088/0264-9381/32/1/015014}

\bibitem[{Musco \& Miller(2013)}]{Musco:2012au}
Musco, I., \& Miller, J.~C. 2013, Class. Quant. Grav., 30, 145009,
  \dodoi{10.1088/0264-9381/30/14/145009}

\bibitem[{Nakama {et~al.}(2017)Nakama, Silk, \& Kamionkowski}]{Nakama:2016gzw}
Nakama, T., Silk, J., \& Kamionkowski, M. 2017, Phys. Rev. D, 95, 043511,
  \dodoi{10.1103/PhysRevD.95.043511}

\bibitem[{Nakamura(2007)}]{Nakamura:2004rm}
Nakamura, K. 2007, Prog. Theor. Phys., 117, 17, \dodoi{10.1143/PTP.117.17}

\bibitem[{Niemeyer \& Jedamzik(1999)}]{Niemeyer:1999ak}
Niemeyer, J.~C., \& Jedamzik, K. 1999, Phys. Rev. D, 59, 124013,
  \dodoi{10.1103/PhysRevD.59.124013}

\bibitem[{Niikura {et~al.}(2019)}]{Niikura:2017zjd}
Niikura, H., {et~al.} 2019, Nature Astron., 3, 524,
  \dodoi{10.1038/s41550-019-0723-1}

\bibitem[{\"Ozsoy {et~al.}(2018)\"Ozsoy, Parameswaran, Tasinato, \&
  Zavala}]{Ozsoy:2018flq}
\"Ozsoy, O., Parameswaran, S., Tasinato, G., \& Zavala, I. 2018, JCAP, 07, 005,
  \dodoi{10.1088/1475-7516/2018/07/005}

\bibitem[{Pi \& Sasaki(2020)}]{Pi:2020otn}
Pi, S., \& Sasaki, M. 2020, JCAP, 09, 037,
  \dodoi{10.1088/1475-7516/2020/09/037}

\bibitem[{Pi {et~al.}(2018)Pi, Zhang, Huang, \& Sasaki}]{Pi:2017gih}
Pi, S., Zhang, Y.-l., Huang, Q.-G., \& Sasaki, M. 2018, JCAP, 05, 042,
  \dodoi{10.1088/1475-7516/2018/05/042}

\bibitem[{{Planck Collaboration} {et~al.}(2020){Planck Collaboration},
  {Aghanim}, {Akrami}, {Ashdown}, {Aumont}, {Baccigalupi}, {Ballardini},
  {Banday}, {Barreiro}, {Bartolo}, {Basak}, {Battye}, {Benabed}, {Bernard},
  {Bersanelli}, {Bielewicz}, {Bock}, {Bond}, {Borrill}, {Bouchet}, {Boulanger},
  {Bucher}, {Burigana}, {Butler}, {Calabrese}, {Cardoso}, {Carron},
  {Challinor}, {Chiang}, {Chluba}, {Colombo}, {Combet}, {Contreras}, {Crill},
  {Cuttaia}, {de Bernardis}, {de Zotti}, {Delabrouille}, {Delouis}, {Di
  Valentino}, {Diego}, {Dor{\'e}}, {Douspis}, {Ducout}, {Dupac}, {Dusini},
  {Efstathiou}, {Elsner}, {En{\ss}lin}, {Eriksen}, {Fantaye}, {Farhang},
  {Fergusson}, {Fernandez-Cobos}, {Finelli}, {Forastieri}, {Frailis},
  {Fraisse}, {Franceschi}, {Frolov}, {Galeotta}, {Galli}, {Ganga},
  {G{\'e}nova-Santos}, {Gerbino}, {Ghosh}, {Gonz{\'a}lez-Nuevo}, {G{\'o}rski},
  {Gratton}, {Gruppuso}, {Gudmundsson}, {Hamann}, {Handley}, {Hansen},
  {Herranz}, {Hildebrandt}, {Hivon}, {Huang}, {Jaffe}, {Jones}, {Karakci},
  {Keih{\"a}nen}, {Keskitalo}, {Kiiveri}, {Kim}, {Kisner}, {Knox},
  {Krachmalnicoff}, {Kunz}, {Kurki-Suonio}, {Lagache}, {Lamarre}, {Lasenby},
  {Lattanzi}, {Lawrence}, {Le Jeune}, {Lemos}, {Lesgourgues}, {Levrier},
  {Lewis}, {Liguori}, {Lilje}, {Lilley}, {Lindholm}, {L{\'o}pez-Caniego},
  {Lubin}, {Ma}, {Mac{\'\i}as-P{\'e}rez}, {Maggio}, {Maino}, {Mandolesi},
  {Mangilli}, {Marcos-Caballero}, {Maris}, {Martin}, {Martinelli},
  {Mart{\'\i}nez-Gonz{\'a}lez}, {Matarrese}, {Mauri}, {McEwen}, {Meinhold},
  {Melchiorri}, {Mennella}, {Migliaccio}, {Millea}, {Mitra},
  {Miville-Desch{\^e}nes}, {Molinari}, {Montier}, {Morgante}, {Moss}, {Natoli},
  {N{\o}rgaard-Nielsen}, {Pagano}, {Paoletti}, {Partridge}, {Patanchon},
  {Peiris}, {Perrotta}, {Pettorino}, {Piacentini}, {Polastri}, {Polenta},
  {Puget}, {Rachen}, {Reinecke}, {Remazeilles}, {Renzi}, {Rocha}, {Rosset},
  {Roudier}, {Rubi{\~n}o-Mart{\'\i}n}, {Ruiz-Granados}, {Salvati}, {Sandri},
  {Savelainen}, {Scott}, {Shellard}, {Sirignano}, {Sirri}, {Spencer},
  {Sunyaev}, {Suur-Uski}, {Tauber}, {Tavagnacco}, {Tenti}, {Toffolatti},
  {Tomasi}, {Trombetti}, {Valenziano}, {Valiviita}, {Van Tent}, {Vibert},
  {Vielva}, {Villa}, {Vittorio}, {Wandelt}, {Wehus}, {White}, {White},
  {Zacchei}, \& {Zonca}}]{Planck2018_parameters}
{Planck Collaboration}, {Aghanim}, N., {Akrami}, Y., {et~al.} 2020, \aap, 641,
  A6, \dodoi{10.1051/0004-6361/201833910}

\bibitem[{Polnarev \& Zembowicz(1991)}]{Polnarev:1988dh}
Polnarev, A., \& Zembowicz, R. 1991, Phys. Rev. D, 43, 1106,
  \dodoi{10.1103/PhysRevD.43.1106}

\bibitem[{Press \& Schechter(1974)}]{Press:1973iz}
Press, W.~H., \& Schechter, P. 1974, Astrophys. J., 187, 425,
  \dodoi{10.1086/152650}

\bibitem[{Ray {et~al.}(2021)Ray, Laha, Mu\~noz, \& Caputo}]{Ray:2021mxu}
Ray, A., Laha, R., Mu\~noz, J.~B., \& Caputo, R. 2021, Phys. Rev. D, 104,
  023516, \dodoi{10.1103/PhysRevD.104.023516}

\bibitem[{Riess {et~al.}(2016)}]{Riess:2016jrr}
Riess, A.~G., {et~al.} 2016, Astrophys. J., 826, 56,
  \dodoi{10.3847/0004-637X/826/1/56}

\bibitem[{Riess {et~al.}(2018)}]{Riess:2018uxu}
---. 2018, Astrophys. J., 855, 136, \dodoi{10.3847/1538-4357/aaadb7}

\bibitem[{{Riess} {et~al.}(2022){Riess}, {Yuan}, {Macri}, {Scolnic}, {Brout},
  {Casertano}, {Jones}, {Murakami}, {Anand}, {Breuval}, {Brink}, {Filippenko},
  {Hoffmann}, {Jha}, {D'arcy Kenworthy}, {Mackenty}, {Stahl}, \&
  {Zheng}}]{Riess2022}
{Riess}, A.~G., {Yuan}, W., {Macri}, L.~M., {et~al.} 2022, \apjl, 934, L7,
  \dodoi{10.3847/2041-8213/ac5c5b}

\bibitem[{{Sch{\"o}neberg} {et~al.}(2022){Sch{\"o}neberg}, {Verde},
  {Gil-Mar{\'\i}n}, \& {Brieden}}]{Verde2022}
{Sch{\"o}neberg}, N., {Verde}, L., {Gil-Mar{\'\i}n}, H., \& {Brieden}, S. 2022,
  \jcap, 2022, 039, \dodoi{10.1088/1475-7516/2022/11/039}

\bibitem[{Serpico(2019)}]{Serpico:2019bcj}
Serpico, P. 2019, PoS, CORFU2018, 094, \dodoi{10.22323/1.347.0094}

\bibitem[{Serpico {et~al.}(2020)Serpico, Poulin, Inman, \&
  Kohri}]{Serpico:2020ehh}
Serpico, P.~D., Poulin, V., Inman, D., \& Kohri, K. 2020, Phys. Rev. Res., 2,
  023204, \dodoi{10.1103/PhysRevResearch.2.023204}

\bibitem[{Takahashi \& Yamada(2019)}]{Takahashi:2019ypv}
Takahashi, F., \& Yamada, M. 2019, JCAP, 07, 001,
  \dodoi{10.1088/1475-7516/2019/07/001}

\bibitem[{Tashiro \& Sugiyama(2013)}]{Tashiro:2012qe}
Tashiro, H., \& Sugiyama, N. 2013, Mon. Not. Roy. Astron. Soc., 435, 3001,
  \dodoi{10.1093/mnras/stt1493}

\bibitem[{Tisserand {et~al.}(2007)}]{EROS-2:2006ryy}
Tisserand, P., {et~al.} 2007, Astron. Astrophys., 469, 387,
  \dodoi{10.1051/0004-6361:20066017}

\bibitem[{Vaskonen \& Veerm\"ae(2021)}]{Vaskonen:2020lbd}
Vaskonen, V., \& Veerm\"ae, H. 2021, Phys. Rev. Lett., 126, 051303,
  \dodoi{10.1103/PhysRevLett.126.051303}

\bibitem[{{Verde} {et~al.}(2019){Verde}, {Treu}, \& {Riess}}]{Verde2019}
{Verde}, L., {Treu}, T., \& {Riess}, A.~G. 2019, Nature Astronomy, 3, 891,
  \dodoi{10.1038/s41550-019-0902-0}

\bibitem[{Wallisch(2018)}]{Wallisch:2018rzj}
Wallisch, B. 2018, PhD thesis, Cambridge U., \dodoi{10.17863/CAM.30368}

\bibitem[{Wang {et~al.}(2018)Wang, Wang, Huang, \& Li}]{Wang:2016ana}
Wang, S., Wang, Y.-F., Huang, Q.-G., \& Li, T. G.~F. 2018, Phys. Rev. Lett.,
  120, 191102, \dodoi{10.1103/PhysRevLett.120.191102}

\bibitem[{{Wang} {et~al.}(2022){Wang}, {Zhang}, {Kimura}, \&
  {Yamaguchi}}]{Wang:2022nml}
{Wang}, X., {Zhang}, Y.-l., {Kimura}, R., \& {Yamaguchi}, M. 2022, arXiv
  e-prints, arXiv:2209.12911.
\newblock \doarXiv{2209.12911}

\bibitem[{Yang(2021)}]{Yang:2021idt}
Yang, Y. 2021, Phys. Rev. D, 104, 063528, \dodoi{10.1103/PhysRevD.104.063528}

\bibitem[{Yokoyama(1998{\natexlab{a}})}]{Yokoyama:1998pt}
Yokoyama, J. 1998{\natexlab{a}}, Phys. Rev. D, 58, 083510,
  \dodoi{10.1103/PhysRevD.58.083510}

\bibitem[{Yokoyama(1998{\natexlab{b}})}]{Yokoyama:1998xd}
---. 1998{\natexlab{b}}, Phys. Rev. D, 58, 107502,
  \dodoi{10.1103/PhysRevD.58.107502}

\bibitem[{Young {et~al.}(2014)Young, Byrnes, \& Sasaki}]{Young:2014ana}
Young, S., Byrnes, C.~T., \& Sasaki, M. 2014, JCAP, 07, 045,
  \dodoi{10.1088/1475-7516/2014/07/045}

\bibitem[{{Yuan} \& {Huang}(2021)}]{Yuan:2021qgz}
{Yuan}, C., \& {Huang}, Q.-G. 2021, iScience, 24, 102860,
  \dodoi{10.1016/j.isci.2021.102860}

\bibitem[{Zhou {et~al.}(2022)Zhou, Lian, Li, Cao, \& Huang}]{Zhou:2021tvp}
Zhou, H., Lian, Y., Li, Z., Cao, S., \& Huang, Z. 2022, Mon. Not. Roy. Astron.
  Soc., 513, 3627, \dodoi{10.1093/mnras/stac915}

\end{thebibliography}
\bibliographystyle{aasjournal}
\end{document}